\documentclass[sigconf]{acmart}
\usepackage{enumitem}
\usepackage{graphicx}
\usepackage{colortbl}
\usepackage{multirow}
\usepackage{xcolor}
\usepackage{tcolorbox} 
\usepackage{booktabs}
\usepackage{subfig}
\usepackage{lipsum}
\usepackage{algorithm}
\usepackage{algorithmic}

\usepackage{pifont} 
\usepackage{xpatch}

\makeatletter
\xpatchcmd{\ps@firstpagestyle}{Manuscript submitted to ACM}{}{\typeout{First patch succeeded}}{\typeout{first patch failed}}
\xpatchcmd{\ps@standardpagestyle}{Manuscript submitted to ACM}{}{\typeout{Second patch succeeded}}{\typeout{Second patch failed}}    \@ACM@manuscriptfalse
\makeatother

\renewcommand\footnotetextcopyrightpermission[1]{}
\setcopyright{none}

\AtBeginDocument{%
  }

\setcopyright{acmlicensed}
\copyrightyear{2018}
\acmYear{2018}
\acmDOI{XXXXXXX.XXXXXXX}
\acmConference[Conference acronym 'XX]{Make sure to enter the correct
  conference title from your rights confirmation email}{June 03--05,
  2018}{Woodstock, NY}

\acmISBN{978-1-4503-XXXX-X/2018/06}
\begin{document}

\title{MGFRec: Towards Reinforced Reasoning Recommendation with Multiple Groundings and Feedback}

\settopmatter{authorsperrow=4}
\author{Shihao Cai}
\email{caishihao@mail.ustc.edu.cn}
\affiliation{%
  \institution{University of Science and Technology of China}
}
\author{Chongming Gao}
\email{chongming.gao@gmail.com}
\affiliation{%
  \institution{University of Science and Technology of China}
}
\author{Haoyan Liu}
\email{liuhaoyan@ustc.edu.cn}
\affiliation{%
  \institution{University of Science and Technology of China}
}
\author{Wentao Shi}
\email{shiwentao123@mail.ustc.edu.cn}
\affiliation{%
  \institution{University of Science and Technology of China}
}
\author{Jianshan Sun}
\email{sunjs9413@hfut.edu.cn}
\affiliation{%
  \institution{Hefei University of Technology}
}
\author{Ruiming Tang}
\email{tangruiming@kuaishou.com}
\affiliation{%
  \institution{Kuaishou Inc.}
}
\author{Fuli Feng}
\email{fulifeng93@gmail.com}
\affiliation{%
  \institution{University of Science and Technology of China}
}
\renewcommand{\shortauthors}{Shihao Cai et al.}

\begin{abstract}
The powerful reasoning and generative capabilities of large language models (LLMs) have inspired researchers to apply them to reasoning-based recommendation tasks, which require in-depth reasoning about user interests and the generation of recommended items.
However, previous reasoning-based recommendation methods have typically performed inference within the language space alone, without incorporating the actual item space.
This has led to over-interpreting user interests and deviating from real items.
Towards this research gap, we propose performing multiple rounds of grounding during inference to help the LLM better understand the actual item space, which could ensure that its reasoning remains aligned with real items.
Furthermore, we introduce a user agent that provides feedback during each grounding step, enabling the LLM to better recognize and adapt to user interests.
Comprehensive experiments conducted on three Amazon review datasets demonstrate the effectiveness of incorporating multiple groundings and feedback.
These findings underscore the critical importance of reasoning within the actual item space, rather than being confined to the language space, for recommendation tasks.

\end{abstract}



\maketitle

\begin{figure}[htbp]
    \centering
    \includegraphics[width=1\linewidth]{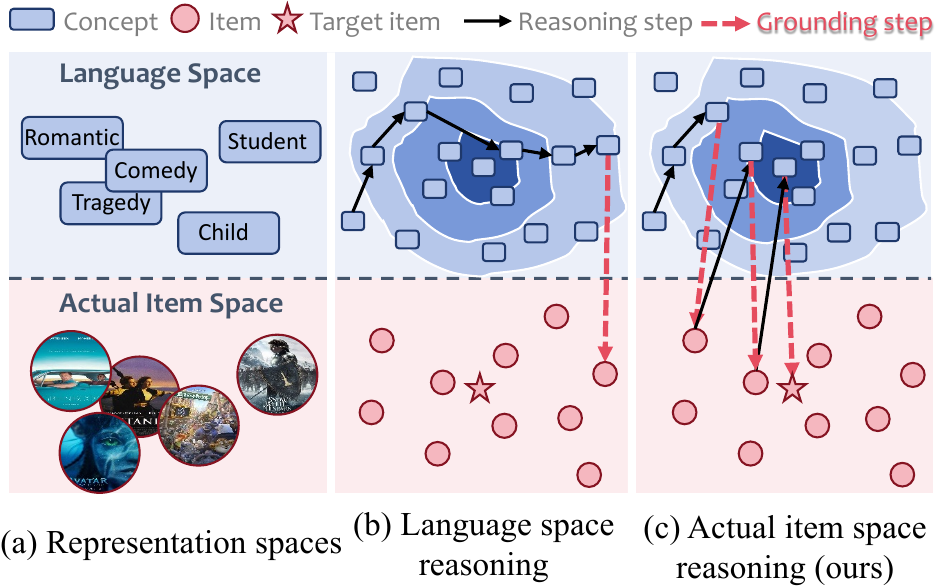}
    \caption{Comparison of reasoning in the language space versus the actual item space. The language space includes all possible constructs, which may lead to recommendations deviating from the target items. In contrast, repeated grounding in the actual item space helps steer the inference towards the target item.}
    \label{fig:comp}
\end{figure}
\section{Introduction}
Large language models (LLMs) have demonstrated impressive generative and reasoning capabilities~\cite{ferrag2025llm,tongyidr,fang2025towards}, inspiring researchers to explore their application in recommendation tasks~\cite{llm_rec_survey,llm_rec_survey_2}.
Early research typically leveraged only one of these abilities.
Some studies focused on generative capabilities, prompting LLMs to directly generate recommended item titles or IDs~\cite{bigrec,llara}.
Other studies concentrated on reasoning abilities, employing LLMs to analyze and infer users' interests in depth~\cite{reason4rec}.
However, effectively harnessing both generative and reasoning abilities simultaneously remains an open challenge.
This challenge has stimulated increased research efforts to enhance the performance of LLMs in generating recommended items through reasoning~\cite{rec_r1,latentr3}.

Among existing approaches, reinforcement learning-based (RL-based) reasoning recommendations have demonstrated significant potential for full item recommendation~\cite{r2ec,latentr3}. 
For example, R$^2$ec \cite{r2ec} first reasons user interests and then recommends items. 
However, as illustrated in Figure~\ref{fig:comp}(b), previous methods typically perform reasoning exclusively in the language space and only map to the actual item space at the final stage. 
This leads to two main issues:  
(1) Excessive reasoning in the language space may cause the model to over-interpret user interests and recommend non-existent items, thereby deviating from the actual item space.
(2) Reasoning limited to the language space results in a lack of feedback signals throughout the reasoning process, relying solely on the final outcome reward, which is inefficient.

Based on this, we argue that it is more appropriate to incorporate the actual item space into the reasoning process, as illustrated in Figure~\ref{fig:comp}(c).
Specifically, for each input, the model engages in a multi-turn process: it first infers the user’s interests, then generates a candidate item title, which is subsequently grounded to the actual item space to retrieve relevant information.
This process repeats until the model determines that no further information is needed, then outputs the recommended item.
The key innovation of this approach lies in extending the single grounding action used in previous methods to multiple grounding actions throughout the reasoning process.
This iterative grounding offers two main advantages: 
(1) By continuously exploring the real item space, the model incrementally narrows the search space during inference, thereby ensuring that the recommended items are not derivative from the actual item space.
(2) Each grounding step provides access to real items, which facilitates the collection of feedback based on actual items, thus aiding in the optimization of recommendations.

To achieve this, we propose \textbf{MGFRec}, a novel RL framework that enables LLMs to perform multiple groundings and obtain feedback. 
MGFRec builds a recommendation agent capable of reasoning in the actual item space.
Specifically, we decompose the generation process of the recommendation agent into three actions: \textbf{think}, \textbf{ground}, and \textbf{answer}. 
During both training and inference, given a sequence of user-item interaction records as input, the recommendation agent first enters a multi-turn procedure: 
(1) It first infers the user's interests within the language space, which may involve one or more entities, and then generates an item title.
(2) This title is then used in the grounding stage to retrieve a list of the most relevant items from the actual item space.
(3) Subsequently, we introduce textual feedback from the user agent to critique the actions of the recommendation agent.
This cycle continues until the recommendation agent determines that no further information is required, after which it proceeds to the answer stage and outputs the recommended items. 
We only employ an outcome-based reward function to evaluate the recommendation results, as the user agent serves as a potential step-level process supervisor.
Finally, to reduce the training cost of RL, we adopt Group Relative Policy Optimization (GRPO)~\cite{shao2024deepseekmath} as the RL algorithm, which eliminates the need for a trainable critic model.

To validate the effectiveness of the proposed framework, we conducted comprehensive experiments on three real-world Amazon review datasets: Books, Movies and TV, and CDs and Vinyl.
The experimental results demonstrate that MGFRec can significantly enhance recommendation performance.
The further in-depth analysis reveals that multiple groundings can reduce the search space and help discover less popular items.

In conclusion, our main contributions are summarized as follows:
\begin{itemize}[topsep=0pt,itemsep=0pt,parsep=0pt,leftmargin=*]
    \item Our work highlights the importance of incorporating actual item space reasoning, rather than limiting reasoning to the language space alone, in LLM-based recommendation tasks.
    \item We propose a novel framework, MGFRec, which extends the single-step grounding process into multi-step grounding and incorporates a user agent to provide feedback.
    \item Extensive experiments validate the effectiveness of the proposed MGFRec approach, underscoring the potential and importance of multiple groundings and feedback in reasoning recommendation.
    
\end{itemize}
\section{Related Work}
In this section, we delve into related studies from three perspectives: reinforcement learning for LLM reasoning, LLM-based recommendation and reasoning recommendation.
\subsection{Reinforcement Learning for LLM Reasoning}
Reinforcement Learning with Human Feedback (RLHF)~\cite{rlhf, rlhf_survey} first introduced reinforcement learning (RL)~\cite{rl} based on human feedback into LLM training. However, it primarily focused on general domains and required the training of a reward model. To reduce computational costs, DPO~\cite{dpo} eliminates the need for a reward model by deriving a closed-form optimal policy from RLHF. 
While early approaches improved the models' mathematical reasoning abilities, the length of their reasoning chains remained limited. DeepSeek-Math~\cite{shao2024deepseekmath} introduced the GRPO algorithm, and DeepSeek-R1~\cite{deepseekr1} further advanced this by removing both critic and reward models, supervising the LLM solely with final rewards. This enabled LLMs to achieve ultra-long reasoning chains.
The success of long reasoning chains in mathematical tasks with models such as GPT-o1~\cite{gpto1} and DeepSeek-R1 has motivated researchers to explore RL-based LLM reasoning in other domains~\cite{searchr1, rec_r1, tabler1, vlmr1}. For example, Search-R1~\cite{searchr1} uses RL to train LLMs for multi-turn interactions, where the model first reasons and then invokes search tools. ReTool~\cite{retool} proposes an automated RL framework that teaches the model when and how to invoke tools based on outcome feedback.
These advancements have spurred research into applying RL-based LLM reasoning in the recommendation domain, with the aim of enhancing the inference of user interests.

\subsection{LLM-based Recommendation}
Due to LLMs' extensive knowledge, researchers have begun exploring their application in the recommendation field~\cite{llm_rec_survey,llm_rec_survey_2,afl,bao2024large}.
Previous work can be roughly categorized into using LLMs for feature enhancement and using LLMs for direct recommendation~\cite{rah}.
For feature enhancement, some researchers enhance feature representation and recommendation models by incorporating LLM-augmented data and embeddings, leveraging LLMs' world knowledge to improve item-user connection learning~\cite{he2023large,hou2023learning, hou2022towards,wang2023enhancing}.
For direct recommendation, some studies leverage LLMs' inherent reasoning abilities for recommendations based on user and interaction data~\cite{reason4rec}, while others fine-tune LLMs on recommendation datasets to enhance domain-specific performance~\cite{collm,sprec,kpo,gao2025process}.
Among them, BIGRec~\cite{bigrec} aligns LLMs from the language space to the actual item space through bi-step grounding. 
What's more, $D^3$~\cite{d3} introduces debiasing techniques during the inference stage to further enhance performance.
In this paper, we focus on enhancing the reasoning capabilities of LLMs to facilitate direct recommendation.

\subsection{Reasoning-Based Recommendation Systems}
\begin{figure*}[htbp]
\centering
\includegraphics[width=0.9\textwidth]{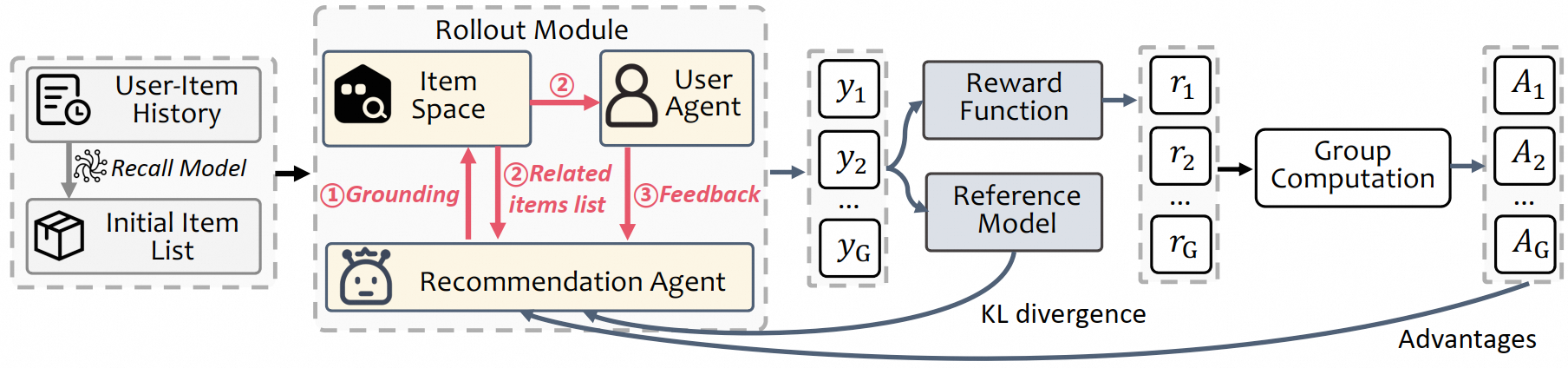}
\caption{Framework of MGFRec.
For each input, MGFRec operates in multiple iterative cycles:
\ding{172} The recommendation agent begins by performing reasoning, followed by grounding operations over the item space.
\ding{173} The grounding operation returns a list of relevant items to both the recommendation agent and the user agent.
\ding{174} The user agent provides feedback to the recommendation agent.
This loop continues until the recommendation agent determines that no further information is needed and outputs the final recommendation result.
}
\label{fig:framework}
\end{figure*}
The powerful reasoning capabilities of LLMs demonstrate their tremendous potential in inferring users’ interests~\cite{liu2025lares, zhang2025slow, tang2025think,wang2024llmrg,zhao2025reason}.
For example, Reason4Rec~\cite{reason4rec} introduces three experts to collaboratively perform multi-step reasoning before making recommendations, effectively enhancing the performance of user rating prediction.
ReaRec~\cite{ReaRec} proposes the first inference-time computing framework for recommender systems, which enhances user representations through implicit multi-step reasoning.
REC-R1~\cite{rec_r1} introduces RL, directly optimizing LLM generation based on feedback from a fixed, black-box recommendation model.
However, most previous works did not focus on leveraging the reasoning capabilities of LLMs themselves to directly perform full item recommendation.
To address this, R$^2$ec~\cite{r2ec} proposes a unified large recommender model, which employs a language modeling head for explicit reasoning generation and a recommendation head for item prediction, thereby enabling full item recommendation.
Unlike R$^2$ec, LatentR$^3$~\cite{latentr3} introduces a novel end-to-end training framework that leverages RL to optimize latent reasoning without relying on any chain-of-thought data.
Most previous work has focused on reasoning within the language space, overlooking the importance of the actual item space in recommendation tasks.
In this paper, we focus on the full item recommendation task and leverage the RL algorithm to enhance the model's explicit reasoning capabilities.

\section{Preliminaries}
\subsection{Representation Spaces}
Building upon the framework proposed in BIGRec~\cite{bigrec}, we define two key spaces in the context of recommendation systems: the language space and the actual item space.
\begin{itemize}[topsep=0pt,itemsep=0pt,parsep=0pt,leftmargin=*]
    \item \textbf{Language Space}. This refers to the set of all possible language sequences that an LLM could generate, such as ``The user enjoys romantic movies.'' While this space represents a broad set of potential outputs, it is not directly usable for recommendations due to its potential to diverge significantly from the actual dataset of items being considered.
    \item \textbf{Actual Item Space}. This space consists of the set of all item titles present within the recommendation dataset. It is both practical and necessary to operate within this space for recommendations, as it directly corresponds to the items available for selection.
\end{itemize}

\subsection{Group Relative Policy Optimization (GRPO)}
Typically, the objective of reinforcement learning is:
\begin{equation}
\begin{aligned}
&\max_{\pi_\theta} \mathbb{E}_{x \sim \mathcal{D}, y \sim \pi_{\theta}(\cdot \mid x)} 
\left[ R(x, y) \right] - \beta \mathbb{D}_{\text{KL}} \left[ \pi_{\theta}(y \mid x) \,||\, \pi_{\text{ref}}(y \mid x) \right],
\label{eq:rl_object_old}
\end{aligned}
\end{equation}
where $\pi_{\theta}$ is the policy model, $\pi_{\text{ref}}$ is the reference model, $R(x,y)$ is the reward function, $\beta$ is hyperparameter and $\mathbb{D}_{\text{KL}}$ is KL-divergence measure.
\( x \) denote input samples drawn from the dataset \( \mathcal{D} \), and \( y \) represent the generated outputs.

Unlike classical RL algorithms such as PPO~\cite{ppo}, for each input \( x \), GRPO~\cite{shao2024deepseekmath} samples a group of responses \( \{ y_1, y_2, \dots, y_G \} \) from the old policy model \( \pi_{\text{old}} \).
Each response $y_i$ receives a scalar reward $R_i$, which will then be used to compute the advantage $A_i$ as follows:
\begin{equation}
\begin{aligned}
\small
 A_i = \frac{R_i - mean(R_1, R_2, \dots, R_G)}{std(R_1, R_2, \dots, R_G)}.
\label{eq:adv}
\end{aligned}
\end{equation}

Subsequently, the GRPO objective function can be formulated as follows:
\begin{equation}
\small
\begin{aligned}
&\mathcal{J}_{GRPO}(\theta) = 
\mathbb{E}_{x \sim \mathcal{D}, \{ y_i \}_{i=1}^{G} \sim \pi_{\text{old}}( \cdot| x)}
\Bigg[
\frac{1}{G} \sum_{i=1}^{G}   
\min \Bigg( 
\frac{\pi_{\theta}(y_{i} | x)}{\pi_{\text{old}}(y_{i} | x)} A_{i}, \\& \text{clip} \Bigg( \frac{\pi_{\theta}(y_{i} | x)}{\pi_{\text{old}}(y_{i} | x)}, 1 - \epsilon, 1 + \epsilon \Bigg) A_{i} 
\Bigg)
- \beta \mathbb{D}_{KL} \left[ \pi_{\theta}(y \mid x) \,||\, \pi_{\text{ref}}(y \mid x) \right]
\Bigg],
\label{eq:grpo_object_old}
\end{aligned}
\end{equation}
where \( \epsilon \) and \( \beta \) are hyperparameters.

\section{Method}
In this section, we first present the overall framework of MGFRec (Section~\S\ref{sec:overall_framework}), followed by detailed descriptions of its key components: (1) multiple groundings (Section~\S\ref{sec:multiple_groundings}), (2) user agent feedback (Section~\S\ref{sec:user_agent_feedback}), and (3) the RL training strategy (Section~\S\ref{sec:rl_based_training}).
\subsection{Overall Framework}
\label{sec:overall_framework}
As illustrated in Figure~\ref{fig:framework}, the input to our framework is user-item interaction history $[I_1, I_2, \dots, I_n]$, and the output is the reasoning-based recommendation text $y$.

At the beginning, we employ a retriever (SASRec~\cite{sasrec}) to recall an initial list of 30 items based on the user's interaction history. 
This step provides the recommendation agent with preliminary information about the actual item space. 
Next, the recommendation agent leverages both the user's interaction history and the initial item list to reason about the user's preferences. 
It conducts multiple rounds of grounding within the actual item space and collects feedback from the user agent.
Ultimately, the agent generates a recommended item title. 
Following BIGRec~\cite{bigrec}, we then rank all candidate items based on their relevance to the generated title to obtain the final recommendation ranking.

\begin{table}[htbp]
\centering
\caption{\textbf{Prompt Example for the user agent}}
\label{tab:user_agent_prompt}
\begin{tabular}{p{\linewidth}}
\toprule
Act as a user agent. \\
\textbf{Record of items you've interacted with:} \{\} \\
Now, you will be provided with an item title and a list of items from the database related to the item. Reflect on whether the item title is appropriate and provide feedback. \\
Important rules: \\
1. Summarize your interests based on your interaction history. \\
2. Provide feedback on the item title in relation to your interests. \\
3. Your feedback may affirm or deny the suitability of the given item title, or offer suggestions for improvement. \\
4. You may incorporate the list of items related to the item title when providing feedback. \\
\textbf{Given item title:} \{\} \\
\textbf{List of items related to the title:} \{\} \\
Output your feedback. \\
\bottomrule
\end{tabular}
\end{table}

\subsection{Multiple Groundings}
\label{sec:multiple_groundings}
For each grounding action, the recommendation agent first generates a textual item title, denoted as $I_T$.
Subsequently, similar to the approach used in BIGRec~\cite{bigrec}, we compute the L2 distance between the embedding of $I_T$ and the embedding of each actual item. The L2 distance is calculated as follows:
\begin{equation}
\begin{aligned}
D_j = || emb(I_j) - emb(I_T) ||
\label{eq:l2_distance}
\end{aligned}
\end{equation}
where $ I_j $ represents the $ j $-th actual item, and $emb()$ denotes the embedding function.

After calculating the L2 distances for all items, we sort the items by these distances and return the titles of the top 10 items, which are considered to be the most closely related to the $I_T$ item. Each grounding action can be formally represented as $f_G(I_T) = \mathcal{L}_R$, where $\mathcal{L}_R$ denotes the list of relevant items.

During both the training and inference phases, the recommendation agent performs multiple rounds of grounding to thoroughly explore the actual item space.
In practice, the agent first performs reasoning to generate candidate item title $I_T$, then attempts grounding. 
If it determines that the items returned by grounding are not suitable, it can re-infer and re-ground as needed. 
This iterative process continues until the recommendation agent determines that no additional item space information is needed and has identified the most suitable items to recommend to the user.

\subsection{User Agent Feedback}
\label{sec:user_agent_feedback}
It is worth noting that the grounding action only allows the recommendation agent to explore the actual item space, but does not provide any feedback on user preferences. Therefore, we introduced a user agent that, after each grounding operation, delivers simulated user feedback. This helps the recommendation agent better understand user preferences.

\definecolor{lightred}{RGB}{255, 204, 204}
\definecolor{lightblue}{RGB}{173,216,230}
\definecolor{lightyellow}{RGB}{255, 255, 204}
\definecolor{lightgreen}{RGB}{144,238,144}
\definecolor{lightorange}{RGB}{255, 200, 100}
\begin{table}[htbp]
\centering
\caption{Template for MGFRec, where the agent can choose between the actions: \{\textit{think}, \textit{ground}, \textit{answer}\}.}
\label{tab:template}
\begin{tabular}{p{\linewidth}}
\toprule
You are a helpful recommendation agent who provides well-reasoned and detailed responses. \\
You must conduct reasoning inside \colorbox{lightred}{\textless think\textgreater{}} and \colorbox{lightred}{\textless /think\textgreater{}} first every time you get new information. \\
After reasoning, if you want to find items in the item database, you can call a grounding engine by using \colorbox{lightblue}{\textless ground\textgreater{}} item title \colorbox{lightblue}{\textless /ground\textgreater{}}. It will return the top relevant items between \colorbox{lightyellow}{\textless item\_list\textgreater{}} and \colorbox{lightyellow}{\textless /item\_list\textgreater{}}, as well as feedback from a user agent between \colorbox{lightgreen}{\textless feedback\textgreater{}} and \colorbox{lightgreen}{\textless /feedback\textgreater{}}. \\
You can use the feedback to conduct further reasoning inside \colorbox{lightred}{\textless think\textgreater{}} and \colorbox{lightred}{\textless /think\textgreater{}}, or you may call the grounding engine again. You may repeat the reasoning and grounding process as many times as needed. \\
If you find that no further external information is needed, you can directly provide one recommended item inside \colorbox{lightorange}{\textless answer\textgreater{}} and \colorbox{lightorange}{\textless /answer\textgreater{}}. \\
\bottomrule
\end{tabular}
\end{table}

Specifically, the user agent is initialized with the user-item interaction history $[I_1, I_2, \dots, I_n]$. It takes as input a grounding item title $I_T$ and the corresponding list of related items $\mathcal{L}_R$. Based on this information, the user agent reasons and generates a simulated textual user feedback $F$. This process can be formally represented as:
$f_U([I_1, I_2, \dots, I_n], I_T, \mathcal{L}_R) = F$.
The specific user agent prompt is presented in Table~\ref{tab:user_agent_prompt}.

\subsection{RL-based Training}
\label{sec:rl_based_training}
Building on the proven effectiveness of GRPO in DeepSeek-R1~\cite{deepseekr1}, we have selected it as the core algorithm for our reinforcement learning training.
\subsubsection{Training Template}
As shown in Table~\ref{tab:template}, this template organizes the recommendation agent's output into three parts for each iterative round: first, a reasoning process, then a grounding action, and finally, the answer. In each round, the recommendation agent starts by reasoning to analyze the user's preferences. Next, it attempts a grounding action to retrieve a relevant list of items and receives feedback from the user agent. If the recommendation agent determines that no additional information is needed, it directly outputs an answer; otherwise, it continues to iterate.

\subsubsection{GRPO with Multiple Groundings and Feedback}
Referring to Search-R1~\cite{searchr1}, we can formulate the RL objective function using a grounding action $f_G$ and a user agent $f_U$ as follows:
\begin{equation}
\begin{aligned}
&\max_{\pi_\theta} \mathbb{E}^{*}_{x \sim \mathcal{D}, y \sim \pi_{\theta}(\cdot \mid x; f_G, f_U)} 
\left[ R(x, y) \right] \\
&- \beta \mathbb{D}_{\text{KL}} \left[ \pi_{\theta}(y \mid x; f_G, f_U) \,||\, \pi_{\text{ref}}(y \mid x; f_G, f_U) \right],
\label{eq:rl_object}
\end{aligned}
\end{equation}
where \( y \) represent the generated outputs interleaved with grounding results and user agent feedback. 

\medskip\noindent\textbf{Objective of GRPO with Multiple Groundings and Feedback}. 
Based on Eq.~\eqref{eq:grpo_object_old} and Eq.~\eqref{eq:rl_object}, we can derive the objective for GRPO with multiple groundings and feedback as follows:
\begin{equation}
\small
\begin{aligned}
&\mathcal{J}_{GRPO}^{*}(\theta) = 
\mathbb{E}_{x \sim \mathcal{D}, \{ y_i \}_{i=1}^{G} \sim \pi_{\text{old}}( \cdot| x; f_G, f_U)}
\Bigg[
\frac{1}{G} \sum_{i=1}^{G}   \\& 
\min \Bigg( 
\frac{\pi_{\theta}(y_{i} | x; f_G, f_U )}{\pi_{\text{old}}(y_{i} | x;  f_G, f_U)} A_{i}, \text{clip} \Bigg( \frac{\pi_{\theta}(y_{i} | x;  f_G, f_U)}{\pi_{\text{old}}(y_{i} | x;  f_G, f_U)}, 1 - \epsilon, 1 + \epsilon \Bigg) A_{i} 
\Bigg)
\\ &- \beta \mathbb{D}_{KL} \left[ \pi_{\theta}(y \mid x; f_G, f_U) \,||\, \pi_{\text{ref}}(y \mid x; f_G, f_U) \right]
\Bigg].
\label{eq:grpo_object}
\end{aligned}
\end{equation}
\medskip\noindent\textbf{Loss Masking for Non-LLM-Generated Tokens}.
In original GRPO, the loss is calculated by all the generated tokens in the whole response.
However, as highlighted in previous work~\cite{ReSearch, r1_searcher, searchr1}, for responses that contain additional tokens not generated by the LLM, simply calculating the loss over all tokens can lead to a biased policy model.
To mitigate the effect, we designate \texttt{\textless item\_list\textgreater{} ... \textless /item\_list\textgreater{}} and \texttt{\textless feedback\textgreater{} ... \textless /feedback\textgreater{}} as special tokens and mask them during training.

\subsubsection{Reward Modeling}
To align the LLM with recommendation objectives, we design a reward function based on the Normalized Discounted Cumulative Gain (NDCG) metric. 
It can be regarded as a special case of NDCG when there is only one relevant item.
Specifically, for each response $y_i$, we obtain the ground truth item's rank $r_i$ among all candidate items, and then compute the recommendation reward function as follows:
\begin{equation}
\begin{aligned}
 R_{\text{rec}, i} = NDCG= \frac{1}{log_2(1 + r_i)}.
\label{eq:reward_rec}
\end{aligned}
\end{equation}

In addition, we also supervise the format of the response $y_i$. 
If $y_i$ does not follow the template provided in Table~\ref{tab:template}, the final reward $R_i$ is set to -0.5. 
Therefore, the function for $R_i$ is as follows:
\begin{equation}
    R_{i} =
\begin{cases}
R_{\text{rec}, i}, & \text{if the format is correct} \\
\text{-0.5}, & \text{if the format is incorrect}
\label{eq:reward_function}
\end{cases}
\end{equation}
\section{Experiments}

In this section, we conduct experiments to answer the following research questions (RQ):

\begin{itemize}[leftmargin=*]
\item \textbf{RQ1}: How does the recommendation performance of MGFRec compare to other baselines?
\item \textbf{RQ2}: What factors contribute to the performance improvements achieved by MGFRec?
\item \textbf{RQ3}: How do the key components of MGFRec affect its overall performance?
\end{itemize}

\subsection{Experimental Setup}

\subsubsection{Datasets}
Similar to previous works on reasoning-based recommendation methods~\cite{r2ec,latentr3}, we conduct our experiments on the Amazon review dataset\footnote{\url{https://amazon-reviews-2023.github.io/index.html}.}, including the categories Books, Movies and TV (Movies), and CDs and Vinyl (CDs).
We adopted the same data processing procedure as R$^2$ec~\cite{r2ec}. 
Specifically, each user's interaction history was chronologically ordered and truncated to the most recent 20 actions, resulting in fixed-length sequences for all subsequent modelling. 
Subsequently, the dataset was divided into training, validation, and test sets in an 8:1:1 ratio. 
The detailed dataset statistics are presented in Table~\ref{tab:dataset_statistics}.

\begin{table}[t]
\centering
\caption{\textbf{Statistics of datasets.}}
\label{tab:dataset_statistics}
\begin{tabular}{lrrr}
\toprule
Dataset& \#Users &\#Item &\#Interaction             \\ 
\toprule
Books&77,348&58,335&203,410 \\
Movies and TV&17,194&19,369&40,041 \\
CDs and Vinyl&7,701&12,024&13,435 \\
\bottomrule
\end{tabular}
\end{table}
\begin{table*}[htbp]
\caption{The recommendation performance of MGFRec compared with various baselines. 
Bold indicates the best performance.}
\label{tab:rec_exp}
\centering
\small
 \renewcommand\tabcolsep{1.5pt} 
 \renewcommand\arraystretch{1.1} 
\begin{tabular}{c|cccccc|cccccc|cccccc}
\toprule
\multirow{2}{*}{Model} & \multicolumn{6}{c|}{Books} & \multicolumn{6}{c|}{Movies and TV} & \multicolumn{6}{c}{CDs and Vinyl}\\
 & H@5 & N@5 & H@10 & N@10 & H@20 & N@20 & H@5 & N@5 & H@10 & N@10 & H@20 & N@20 & H@5 & N@5 & H@10 & N@10 & H@20 & N@20\\
\toprule
SASRec&0.0095&0.0065&0.0143&0.0080&0.0221&0.0099  &0.0075&0.0051&0.0088&0.0055&0.0110&0.0060    & 0.0030&0.0023&0.0038&0.0025&0.0045&0.0027\\
GRU4Rec&0.0019&0.0015&0.0026&0.0017&0.0044&0.0022   &0.0018&0.0012&0.0035&0.0017&0.0053&0.0022    & 0.0030&0.0028&0.0045&0.0032&0.0076&0.0040\\
Caser&0.0017&0.0014&0.0028&0.0018&0.0049&0.0023   &0.0044&0.0033&0.0053&0.0036&0.0079&0.0042     &     0.0015&0.0011&0.0038&0.0018&0.0068&0.0026\\
\midrule
BIGRec&0.0166&0.0125&0.0323&0.0172&0.0460&0.0187   &0.0196&0.0147&0.0284&0.0176&0.0415&0.0194     &  0.0074&0.0052&0.0089&0.0056&0.0149&0.0071\\
$D^3$&0.0192&0.0143&0.0356&0.0195&0.0497&0.0214    & 0.0231&0.0152&0.0326&0.0183&0.0459&0.0201     &  0.0082&0.0064&0.0178&0.0082&0.0203&0.0094\\
\midrule
LatentR$^3$&0.0224&0.0163&0.0405&0.0221&0.0577&0.0254     &0.0273&0.0159&0.0371&0.0189&0.0488&0.0213      & 0.0089&0.0067&0.0186&0.0087&0.0212&0.0099\\
R$^2$ec &0.0430&0.0267&0.0609&0.0331&0.0769&0.0376       & 0.0284&0.0184&\textbf{0.0433}&0.0232&0.0545&0.0246      & 0.0094&0.0073&0.0192&0.0094&0.0226&0.0108\\
\midrule
MGFRec& \textbf{0.0468}&\textbf{0.0334}&\textbf{0.0625}&\textbf{0.0380}&\textbf{0.0801}&\textbf{0.0421}      &\textbf{0.0301}&\textbf{0.0207}&0.0423&\textbf{0.0247}&\textbf{0.0577}&\textbf{0.0286}        &   \textbf{0.0098}&\textbf{0.0078}&\textbf{0.0197}&\textbf{0.0098}&\textbf{0.0234}&\textbf{0.0113}\\

\bottomrule
\end{tabular}
\end{table*}

\subsubsection{Evaluation Setting}
We adopt the full ranking recommendation setting, where the model is required to rank all candidate items. Subsequently, we evaluate the effectiveness of top-N recommendations using Hit Ratio (H@K) and Normalized Discounted Cumulative Gain (N@K), with K set to 5, 10, and 20.

\subsubsection{Baselines}
We have chosen a variety of models as the baselines.
These models can be broadly categorized into traditional recommendation models, non-reasoning LLM-based models, and reasoning LLM-based models.

The traditional recommendation models are as follows:
\begin{itemize}[leftmargin=*]
\item \textit{SASRec}~\cite{sasrec} is an attention-based sequential recommendation model that excels at capturing long-range semantic dependencies across both sparse and dense datasets.
\item \textit{GRU4Rec}~\cite{gru4rec} leverages RNN, offering a relatively simple yet highly efficient approach to sequential recommendation.
\item \textit{Caser}~\cite{caser} models users’ historical behavior sequences as ``images,'' and employs CNN to effectively extract sequential features.
\end{itemize}
The non-reasoning LLM-based models are as follows:
\begin{itemize}[leftmargin=*]
\item \textit{BIGRec}~\cite{bigrec} is a representative LLM-based generative recommendation method
that fine-tunes LLMs to generate next-item predictions, with specific designs to support full ranking.
\item  \textit{D$^3$}~\cite{d3}  follows a similar fine-tuning process to BIGRec but differs in its inference strategy. Specifically, it introduces debiasing techniques during inference to enhance the quality of generated recommendations.
\end{itemize}
The reasoning LLM-based models are as follows:
\begin{itemize}[leftmargin=*]
\item \textit{R$^2$ec}~\cite{r2ec} enhances the recommendation capabilities of LLMs through explicit reasoning, and it also employs a recommendation head for item prediction.
\item  \textit{LatentR$^3$}~\cite{latentr3} employs implicit reasoning, which reduces computational overhead during training and introduces continuous reward signals for more efficient learning.
\end{itemize}

\subsubsection{Implementation Details}
Our experiments are conducted on eight NVIDIA A40 GPUs. 
We employ the Qwen2.5-1.5B-Instruction \cite{qwen2.5} model as the backbone.
Additionally, we use gte-Qwen2-1.5B-instruct~\cite{qwen2embedding} to encode item titles into embeddings. 
The model is trained for one epoch with a learning rate of 1e-6. 
The global batch size is fixed at 128. We set the maximum number of groundings to 6, with each grounding action returning 10 items. For GRPO training, the rollout number is set to 6.
For traditional models, we strictly follow the implementation in~\cite{frame_seq_rec}, using a learning rate of 0.001, an embedding dimension of 64, and a batch size of 256. 
What's more, we also perform a grid search over the values $[1e-3, 1e-4, 1e-5, 1e-6, 1e-7]$ to determine the optimal coefficient for L2 regularization.
For LLM-based models, we strictly followed the original paper’s code for replication, selecting the checkpoint with the best performance on the validation set. 
For the user agent, due to the high frequency of requests during the training process, we employ the ``gpt-4.1-nano-2025-04-14''~\cite{gpt4_1} API.

\subsection{Effectiveness of MGFRec (RQ1)}
In this section, we compare MGFRec with various traditional and LLM-based recommendation models to evaluate the effectiveness of MGFRec. The experimental results are presented in Table~\ref{tab:rec_exp}.

Based on the results in Table~\ref{tab:rec_exp}, we can observe that:
(1) MGFRec outperforms all other baselines in the vast majority of scenarios, thereby demonstrating the effectiveness of incorporating multiple groundings and user agent feedback.
(2) All reasoning models outperform non-reasoning models, which further demonstrates the importance of reasoning in inferring user interests more deeply in recommendation tasks.
(3) MGFRec outperforms other reasoning models, demonstrating that reasoning in the actual item space is superior to reasoning in the language space.
(4) Although MGFRec's H@10 performance on the Movies dataset is lower than that of R$^2$ec, the decrease is relatively small (about 2.3\%), which appears to be acceptable.

\subsection{In-depth Analysis (RQ2)}
In this section, we explore the factors contributing to the performance enhancements observed in MGFRec.
We conduct experimental analyses from three perspectives: sample difficulty (Section~\S\ref{sec:popularity}), the item search space (Section~\S\ref{sec:search_space}) and the training curves (Section~\S\ref{sec:training_curve}). 
Additionally, we present a case study (Section~\S\ref{sec:case_study}) to illustrate how multiple groundings and feedback contribute to the model’s effectiveness.

\subsubsection{Sample Difficulty}
\label{sec:popularity}
\begin{figure}[htbp]
\centering
\includegraphics[width=0.7\columnwidth]{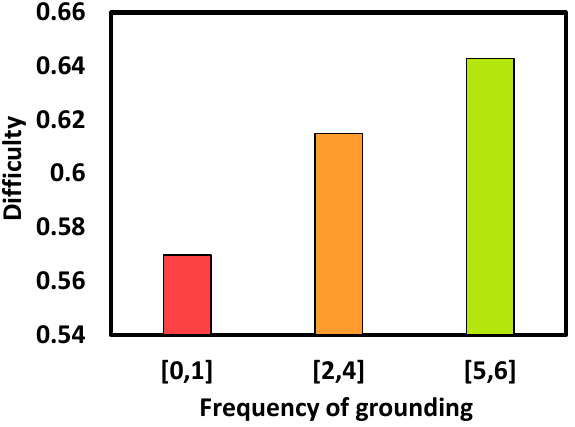}
\caption{Average sample difficulty on the Movies validation set under different grounding frequencies. The difficulty is defined as the inverse of the popularity of the ground truth item for each sample.
}
\label{fig:pop_exp}
\end{figure}
\begin{figure*}[htbp]
\centering
\includegraphics[width=0.85\textwidth]{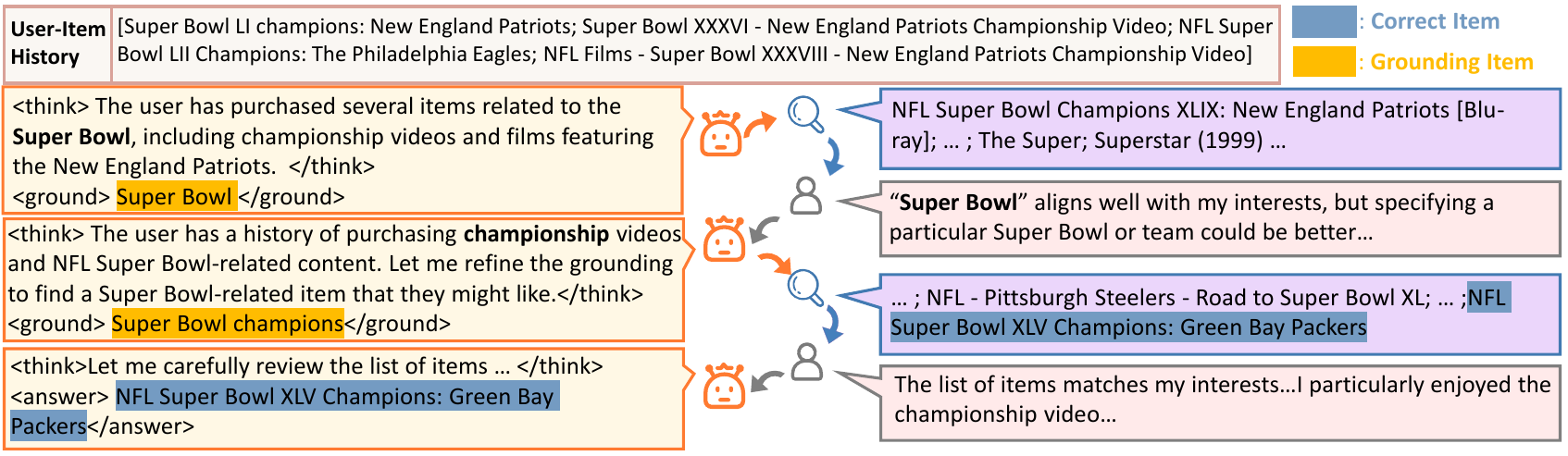}
\caption{The example of how the recommendation agent finds the correct item through multiple groundings and feedback.
}
\label{fig:case_study}
\end{figure*}
\begin{figure}[htbp]
\centering
\includegraphics[width=0.7\columnwidth]{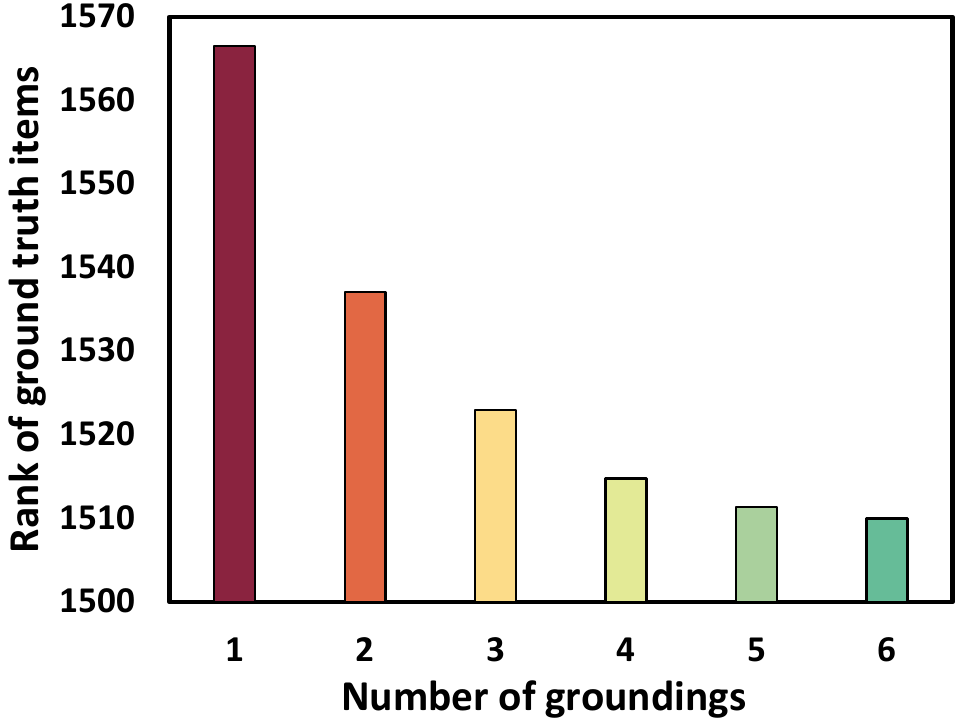}

\caption{Average rank of ground truth items under different maximum grounding numbers on the Movies validation set.
}
\label{fig:item_space}
\end{figure}
In this section, we examine whether more challenging samples require the model to ground more frequently in order to identify ground truth items. 
To investigate this, we conducted experiments on the Movies validation set.
Considering that less popular items are harder to find, for each sample, we use the inverse of the ground-truth item's popularity as a measure of difficulty, where item popularity is determined by the number of user interactions with that item.
We then divided the frequency of grounding into three intervals: low ([0, 1]), medium ([2, 4]), and high ([5, 6]).
Finally, we calculated the average difficulty of samples corresponding to each grounding frequency interval.
The experimental results are presented in Figure~\ref{fig:pop_exp}.

According to the experimental results shown in Figure~\ref{fig:pop_exp}, we can observe a clear trend: as the grounding frequency increases, the average difficulty of the samples also becomes higher.
This reflects the following conclusions: 
(1) Multiple grounding actions help less popular items receive greater exposure, allowing items that cannot be found by a single grounding step to be recommended; 
(2) The LLM learns during training to use multiple grounding actions to progressively identify more suitable items, further validating the effectiveness of multiple grounding.

\subsubsection{Analysis of the Search Space}
\label{sec:search_space}

In this section, we examine whether the model’s search space for relevant items narrows as the number of grounding actions increases during inference. We conducted experiments on the Movies validation set, focusing on samples where the model's predictions were relatively accurate (i.e., the ground truth item ranked within the top 4096). 
During model inference, we constrained the maximum number of groundings and extracted the grounding item title $I_T$ for each sample.
We then determined the rank of the ground truth item based on the procedure described in Section~\S\ref{sec:multiple_groundings}. Finally, we computed the average rank of the ground truth item across different maximum grounding limits. The results are presented in Figure~\ref{fig:item_space}.

According to the experimental results in Figure~\ref{fig:item_space}, we observe that as the number of grounding steps increases, the average rank of ground truth items gradually decreases. This reveals the following insights:
(1) Multiple grounding actions help the ground truth item to progressively stand out within the entire item space, making it easier for the model to identify the ground truth, i.e., the search space is gradually reduced.
(2) During training, the model learns a search paradigm that transitions from exploring the full item space to progressively narrowing down to a smaller space, thus enabling the model to locate user-preferred items step by step through iterative grounding.

\subsubsection{Training Curves}
\label{sec:training_curve}
\begin{figure}[t]
    \centering
    \subfloat[]{\includegraphics[width=0.45\linewidth]{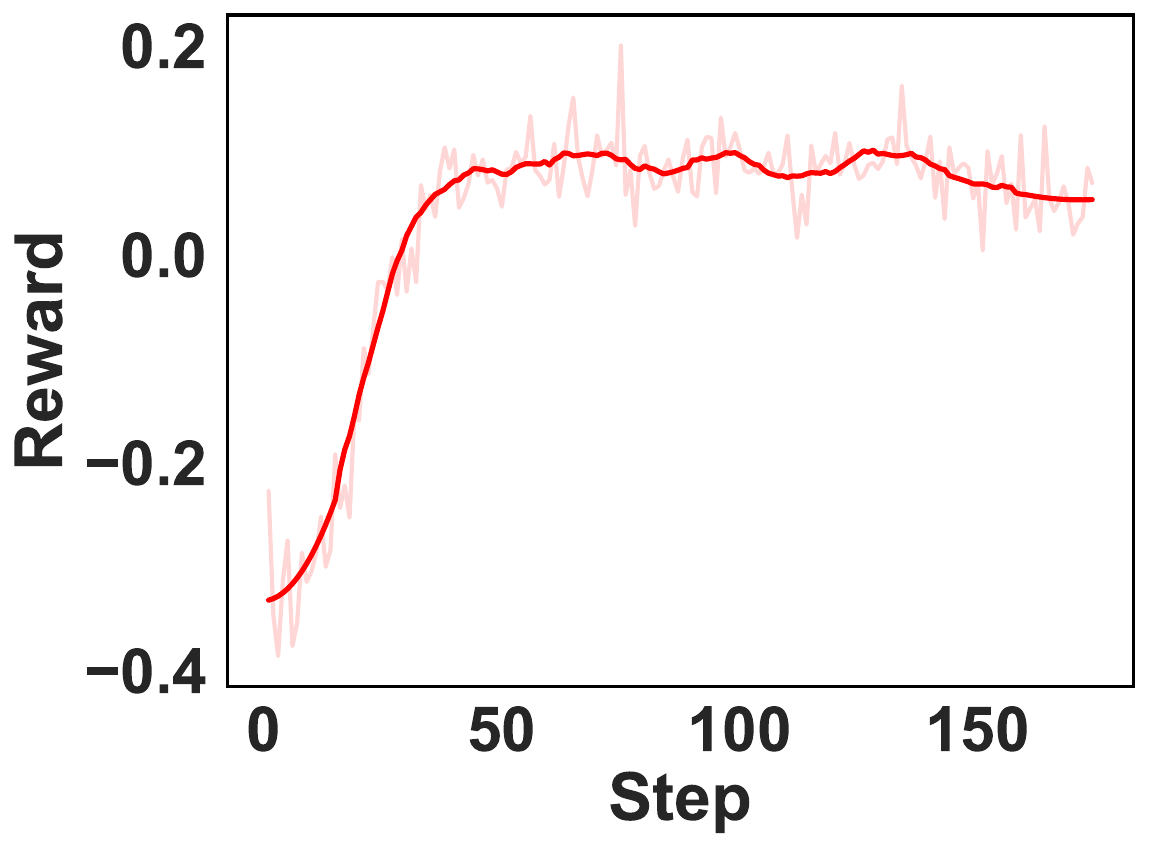}\label{fig:reward}}
    \subfloat[]{\includegraphics[width=0.45\linewidth]{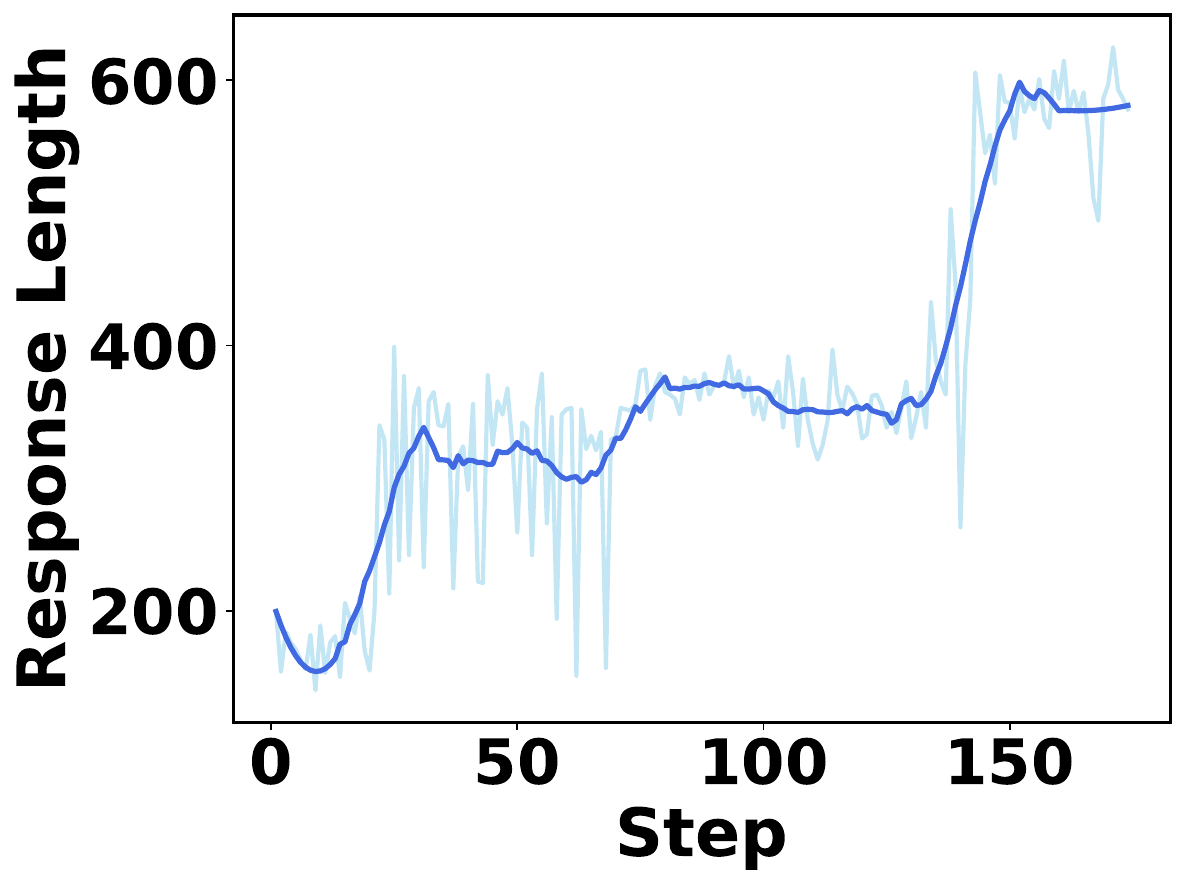}\label{fig:response}}
    \caption{(a) The reward curve on the Movies training set during the training process.
    (b) The response length curve on the Movies training set during the training process.
    }
    \label{fig:reward_response}
\end{figure}

To investigate the rewards obtained by the recommendation agent and the response lengths during training, we plotted the corresponding curves on the Movies training set, as shown in Figure~\ref{fig:reward_response}.

Based on the curves in Figure~\ref{fig:reward_response}, we can draw the following conclusions:
(1) The model quickly learns the response format, leading to a rapid increase in reward. It then gradually learns recommendation capabilities, during which the reward oscillates and eventually converges.
(2) The length of the model’s responses gradually increases, indicating that longer chains of reasoning contribute positively to recommendation performance.
\subsubsection{Case Study}
\label{sec:case_study}

To demonstrate the effectiveness of multiple groundings and feedback, we present an example in Figure~\ref{fig:case_study}.
Initially, the recommendation agent analyzes the user-item interaction history and finds that the user enjoys ``Super Bowl.''
The grounding engine then returns a list of related items, but the user agent provides feedback, suggesting that the results should be more specific. 
In the second round, the recommendation agent further analyzes the user's preferences, identifying a stronger interest in championship videos, and refines the grounding item title accordingly. 
This time, the user agent is satisfied with the list of items returned by the grounding engine. 
Finally, the recommendation agent successfully identifies the correct ground truth item from the list.

In conclusion, during the response generation phase, the recommendation agent dynamically infers user interests and acquires information about the actual item space through multiple rounds of grounding. 
Meanwhile, the user agent provides feedback based on this information, helping the recommendation agent better analyze and optimize the recommendation results. 
This collaborative process ultimately leads to recommendations that not only align with user preferences but also closely reflect the real item space.

\subsection{Impact of Key Components (RQ3)}
In this section, we investigate the impact of various components, including ablation studies (Section~\S\ref{sec:ablation}) and hyperparameter experiments (Section~\S\ref{sec:hyperparameter}).
\begin{table}[htbp]
\caption{Comparison of performance under different settings. Bold results indicate the best results.}
\label{tab:ablation_exp}
\centering
\small
 \renewcommand\tabcolsep{2.4pt} 
 \renewcommand\arraystretch{1} 
\begin{tabular}{l|cccccc}
\toprule
\multirow{2}{*}{Method} & \multicolumn{6}{c}{Movies and TV} \\
 & H@5 & N@5 & H@10 & N@10 & H@20 & N@20 \\
 \toprule
 MGFRec &\textbf{0.0313}&\textbf{0.0214}&\textbf{0.0436}&\textbf{0.0253}&\textbf{0.0590}&\textbf{0.0298} \\
\midrule
w/o Multi Grounding    &0.0258&0.0156&0.0351&0.0184&0.0466&0.0207  \\
w/o Agent Feedback &  0.0281 & 0.0179 & 0.0405&0.0216& 0.0533& 0.0228\\
w/o Recall Model &  0.0293&0.0192&0.0419&0.0235&0.0560&0.0264\\
\bottomrule

\end{tabular}
\end{table}
\subsubsection{Ablation Studies}
\label{sec:ablation}
In this section, we explore the impact of three key modules on model performance: multiple grounding, agent feedback, and the recall model.
We conducted experiments on the Movies validation set, and the results are shown in Table~\ref{tab:ablation_exp}.

Based on the experimental results in Table 1, we draw the following conclusions:
(1) The performance of MGFRec declines whenever any one of its modules is removed, demonstrating the effectiveness of all three components.
(2) The multiple grounding module has the greatest impact on performance, further confirming that the performance gains of MGFRec are primarily attributed to acquiring actual item space information through multiple rounds of grounding during the response generation process.
(3) The recall model has the least impact on performance. In practice, the recall model merely provides the recommendation agent with an initial list of recommended items, which the recommendation agent can either reference or supplement by obtaining a new list through its own grounding. Therefore, the recall model only serves to provide initial information about the item space and is not a critical core component of MGFRec.

\begin{figure}[htbp]
\centering
\includegraphics[width=0.8\columnwidth]{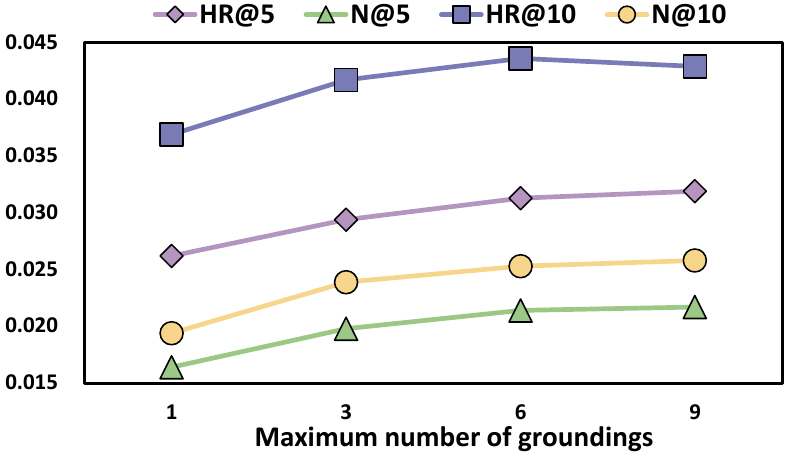}

\caption{Performance on the Movies validation set under different maximum numbers of groundings in training settings.
}
\label{fig:grounding_num}
\end{figure}
\subsubsection{Hyperparameter}
\label{sec:hyperparameter}
In this section, we examine how model performance is affected by two hyperparameters: the maximum number of groundings and the number of returned items.

\textbf{Maximum number of groundings}.
\begin{figure}[htbp]
\centering
\includegraphics[width=0.8\columnwidth]{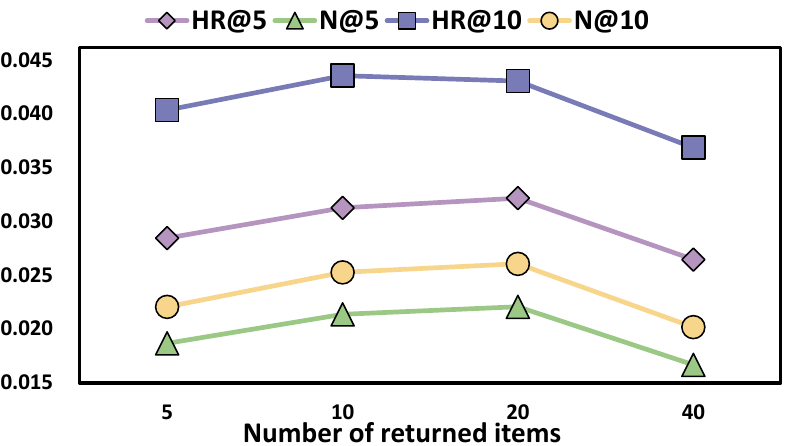}

\caption{Performance on the Movies validation set under different numbers of returned items in training settings.
}
\label{fig:return_num}
\end{figure}
We fix the number of items returned per grounding to 10, and vary the maximum number of groundings in the training settings among $[1, 3, 6, 9]$.
The performance evaluation results on the Movies validation set are shown in Figure~\ref{fig:grounding_num}. 
Based on the results, we can observe the following: (1) In general, a larger maximum number of groundings leads to better performance, highlighting the importance of multiple groundings; 
(2) However, as the maximum number increases, the performance improvement gradually plateaus. Considering that increasing the maximum number also results in longer response generation times, it is important to strike a balance between performance and computational cost.

\textbf{Number of returned items}.
We fixed the maximum number of groundings at 6, and varied the number of items returned per grounding among $[5, 10, 20, 40]$.
The performance evaluation results on the Movies validation set are shown in Figure~\ref{fig:return_num}.
Based on the experimental results, we observe that returning either too few or too many items leads to a decline in performance.
This is because when the number of returned items is too small, the LLM cannot adequately perceive the actual item space; conversely, when too many items are returned, excessive redundant items occupy most of the LLM’s context.

\section{Conclusion and Future Work}
In this study, we propose a novel reinforced framework called MGFRec, which transforms single grounding actions into multiple grounding actions and incorporates user agent feedback.
Through multiple rounds of grounding, our framework shifts the paradigm from reasoning in the language space to reasoning in the actual item space, which enables LLMs to better perceive the actual item space and more effectively capture user feedback.
Extensive experimental results demonstrate the effectiveness of MGFRec and highlight the importance of reasoning in the actual item space for recommendation tasks.

In the future, we will explore more efficient grounding and feedback paradigms.
While our current approach grounds recommendations using item titles, it would be valuable to incorporate additional information such as item genres, styles, and authors. 
Leveraging these attributes could enable the recommendation agent to perform more personalized and flexible grounding.
Moreover, our current user agent is relatively simple. Developing a user agent that more accurately simulates real user behavior to provide higher-quality feedback will be the focus of our future work.
Finally, in the inference phase, MGFRec can be regarded as a special form of multi-turn recommender-user interaction, which demonstrates the potential of applying the MGFRec framework to conversational recommendation scenarios.

\bibliographystyle{ACM-Reference-Format}
\bibliography{sample-base}

\appendix

\end{document}